\documentclass[fleqn,twoside]{article}
\usepackage{espcrc2}

\usepackage{epsfig}

\usepackage[figuresright]{rotating}

\newcommand{\AmS}{{\protect\the\textfont2
  A\kern-.1667em\lower.5ex\hbox{M}\kern-.125emS}}

\title{Study of D-mesons using hadronic decay channels with the ALICE detector}

\author{Renu Bala\address[MCSD] {Universita and INFN, Torino} for the ALICE Collaboration %
        }
       
\begin{document}

\begin{abstract}

At LHC energy, heavy quarks will be abundantly produced and the design
of the ALICE Experiment will allow us to study their production using
several channels. We investigate the feasibility of the study of D
mesons reconstructed in their exclusive hadronic decay channel. After
reviewing the  ALICE potential for such studies, we will present some
results for the two more promising decay channels i.e $D^{0}
\rightarrow K^{-} \pi^{+}$ and $D^{+} \rightarrow K^{-} \pi^{+}
\pi^{+}$ obtained with 7 TeV pp data and 5.5 A TeV Pb-Pb Monte Carlo data .

\vspace{1pc}
\end{abstract}

\maketitle
\section{Motivation}
 ALICE \cite{JINST} is one of the six experiments running at the CERN Large Hadron Collider (LHC). 
It is conceived for the study of heavy-ion collisions up to an energy of  5.5 A TeV and in particular
to explore the QGP phase transition.  Since charm and bottom quarks have large mass, they are produced
 almost exclusively in the initial parton--parton
interactions in the heavy-ion collisions. The time scale for a
$c \bar{c}$ pair production is $\sim \hbar $ / (2 $m_{Q} c^{2}$)
$\approx$ (0.2 GeV fm $c^{-1}$)/ (2.4 GeV) $\approx$ 0.1 fm/c, which
is much smaller than the expected lifetime of the Quark Gluon Plasma
$\sim$ 10 fm/c. Thus heavy quarks are expected to provide information
about the hottest initial phase. The measurement of D-mesons can be
used to extract the charm production cross section. The measurement of
charm cross section in both pp and AA collisions is useful to evaluate
the scaling mechanisms which govern the charm production from pp to AA
collisions. Several nuclear effect can break the binary scaling
estimated on a geometrical basis with the Glauber model \cite{Glaub}. They are
divided into two classes: initial and final state effects. The former,
such as nuclear shadowing, affect heavy quark production by modifying
the parton distribution functions in the nucleus. Initial state
efffects can be studied by comparing proton-proton and proton-nucleus collisions. The later can be
due to the interaction of the partons in the medium. Partonic energy
loss in the medium is the main example of such  an effect.
\section{The ALICE detector}
The ALICE detector  consists of two parts:  a central barrel, which
includes the Inner Tracker System (ITS), the Time-Projection-Chamber
(TPC),  the Transition Radiation Detector(TRD), the Time-Of-Flight system
(TOF),  Electro Magnetic Calorimeter (EMCAL) and High Momentum Particle
Identification detector (HMPID) all
with the full acceptance of $|\eta| <0.9$ and the forward Muon Spectrometer, 
covering the pseudorapidity range between 2.4 and 4. For the  present study we used the information from the following
detector sub-system:
The two inner detectors, the $\bf{ITS}$ and the $\bf{TPC}$, allow the reconstruction of 
charged particle tracks with very good impact parameter and momentum resolution due to their
high granularity and provide particle identification via dE/dx measurement .
The ITS\cite{ITS}, in particular,  is a key detector for open heavy flavour studies because it
allows to measure the track impact parameter (i.e. the distance of
closest approach of the track to the primary vertex) with a resolution
better than 80 $\mu$m for $p_{t} > $ 1.0 GeV/c  thus providing
the capability to detect the secondary vertices originating from
heavy-flavour decays. {\bf TOF} provides  particle identification by
time of flight measurement and is used for the K/$\pi$ separation
below  2 GeV/c 

\section{Charm Reconstruction in the hadronic decay channels} 
\begin{figure}[ht]
\includegraphics[width=16pc,height=10pc]{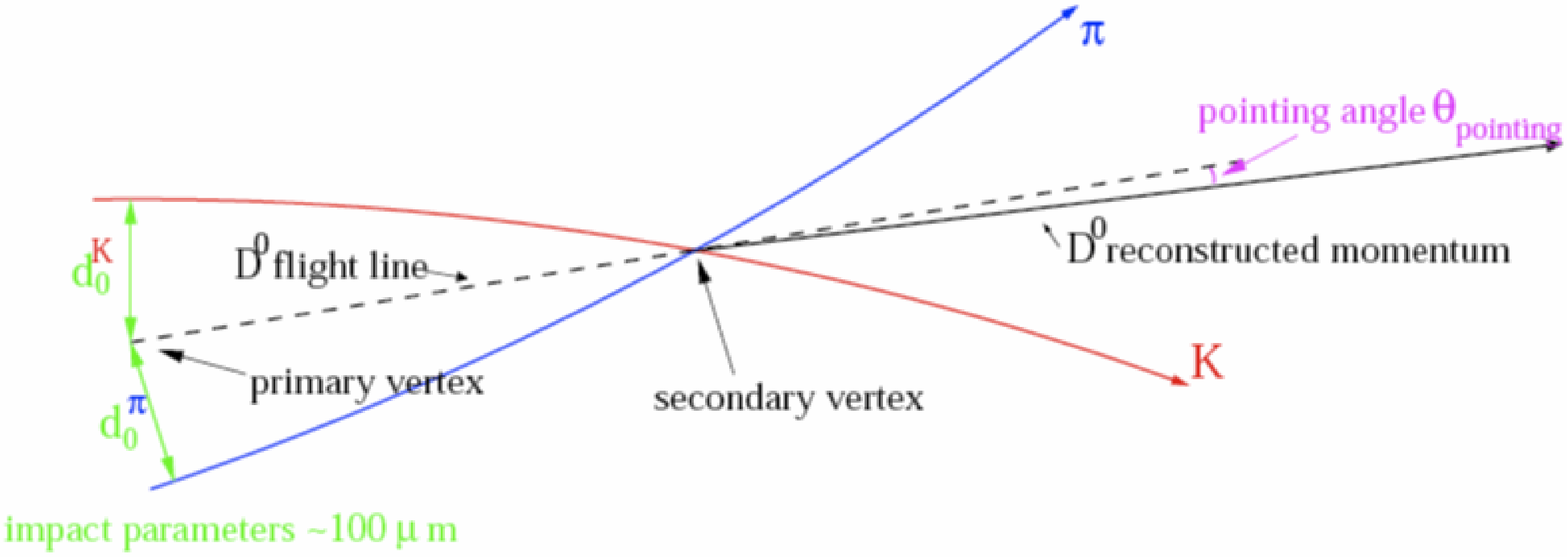}
\includegraphics[width=16pc,height=10pc]{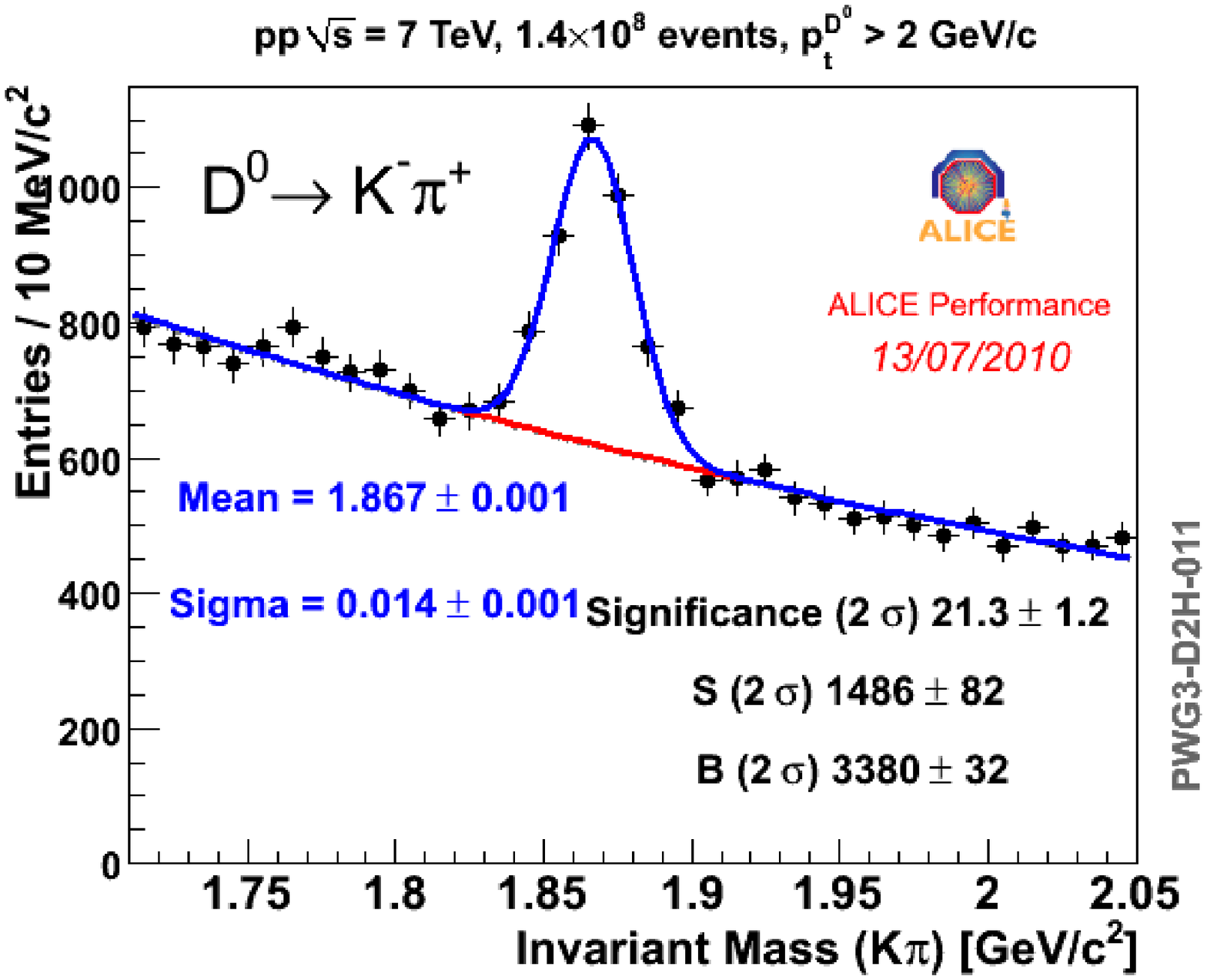}
\caption{Upper Panel: Sketch of $D^{0} \rightarrow K^{-} \pi^{+}$. Lower Panel: Invariant mass spectra for $p_{t}>$ 2 GeV/c}
\label{d0dec}
\end{figure}

An intensive simulation study of  D mesons from hadronic decays has been already done using the
decay channels $D^{0} \rightarrow K^{-} \pi^{+}$\cite{dainese,PPR}, $D^{+} \rightarrow K^{-} \pi^{+} \pi^{+}$ \cite{bruna} and
$D^{*+} \rightarrow D^{0} \pi^{+}_{s}$, $D^{0} \rightarrow K^{-}\pi^{+}\pi^{-}\pi^{+}$ and $D_{s}^{+} \rightarrow K^{+} K^{-} \pi^{+}$ \cite{sergey} showing that ALICE has an excellent capability to carry out such studies. 
The Reconstruction of heavy-flavoured mesons is  a challenging task as these are  rare signals and are having 
large combinatorial background. The detection strategy to cope with
such large combinatorial background  is based on the selection of displaced-vertex topologies i.e
the identification of single track and secondary vertices that are displaced from
the interaction vertex. A good alignment between reconstructed D
momentum and flight line is required.  After the reconstruction,  an invariant-mass 
analysis is used to extract the raw signal yield, which is then corrected for selection and 
reconstruction efficiency and for detector acceptance. 
Here we will briefly explain the selection strategy
for two decay channel $D^{0}\rightarrow K^{-} \pi^{+}$ and
$D^{+}\rightarrow K^{-} \pi^{+} \pi^{+}$ and will show the very first
results obtained from a sample of 1.4 $\times 10^{8}$ minimum bias proton-proton events at a
centre of mass energy  $\sqrt{s}$ =7 TeV \cite{mult} collected during April-May
2010 with the ALICE detector at LHC.
{\bf $D^{0} \rightarrow K^{-} \pi^{+}$}: The main feature of this decay
topology (shown in the upper panel of the fig \ref{d0dec}) is the
large impact parameter of daughter tracks which is of order 100
$\mu$m. Two main variables are used to separate the signal from the
combinatorial background of opposite sign track pairs: the product of the
impact parameters of the two tracks ($d_{0}^{K} \times d_{0}^{\pi}$) and the cosine of the pointing
angle ($\theta_{point}$). The tuning of the cuts has been done in the
present study for each separate $D^{0}$ transverse momentum bin. With
$10^{9}$ p-p events\footnote{Number of events expected to collect in 2010},  we expect a $p_{t}$ integrated significance of 50 and larger than 10 for  $p_{t}$ up to $\sim$ 10 GeV/c.

$D^{+} \rightarrow K^{-} \pi^{+} \pi^{+}$: With respect to the decay $D^{0} \rightarrow K^{-} \pi^{+}$, this topology is affected
by higher combinatorial background but  the longer decay length of the
$D^{+}$ meson ($c \tau \approx 311 \mu m$ compared to that of $D^{0}$
meson 123 $\mu$m ) can be regarded as an advantage for the $D^{+}$
reconstruction since a more displaced secondary vertex should help in
the separation of signal from the background.
 The main variables to separate the signal from  combinatorial
 background of the three charged track combinations are: the distance
 between the primary and secondary vertex and the cosine of pointing
 angle ($\theta_{point}$). If the found vertex really corresponds to the  $D^+$ vertex, then $\theta_{point} \approx$ 0 and Cos$\theta_{point} \approx$ 1.
 \begin{figure*}[!ht]
   \includegraphics[width=18pc,height=10pc]{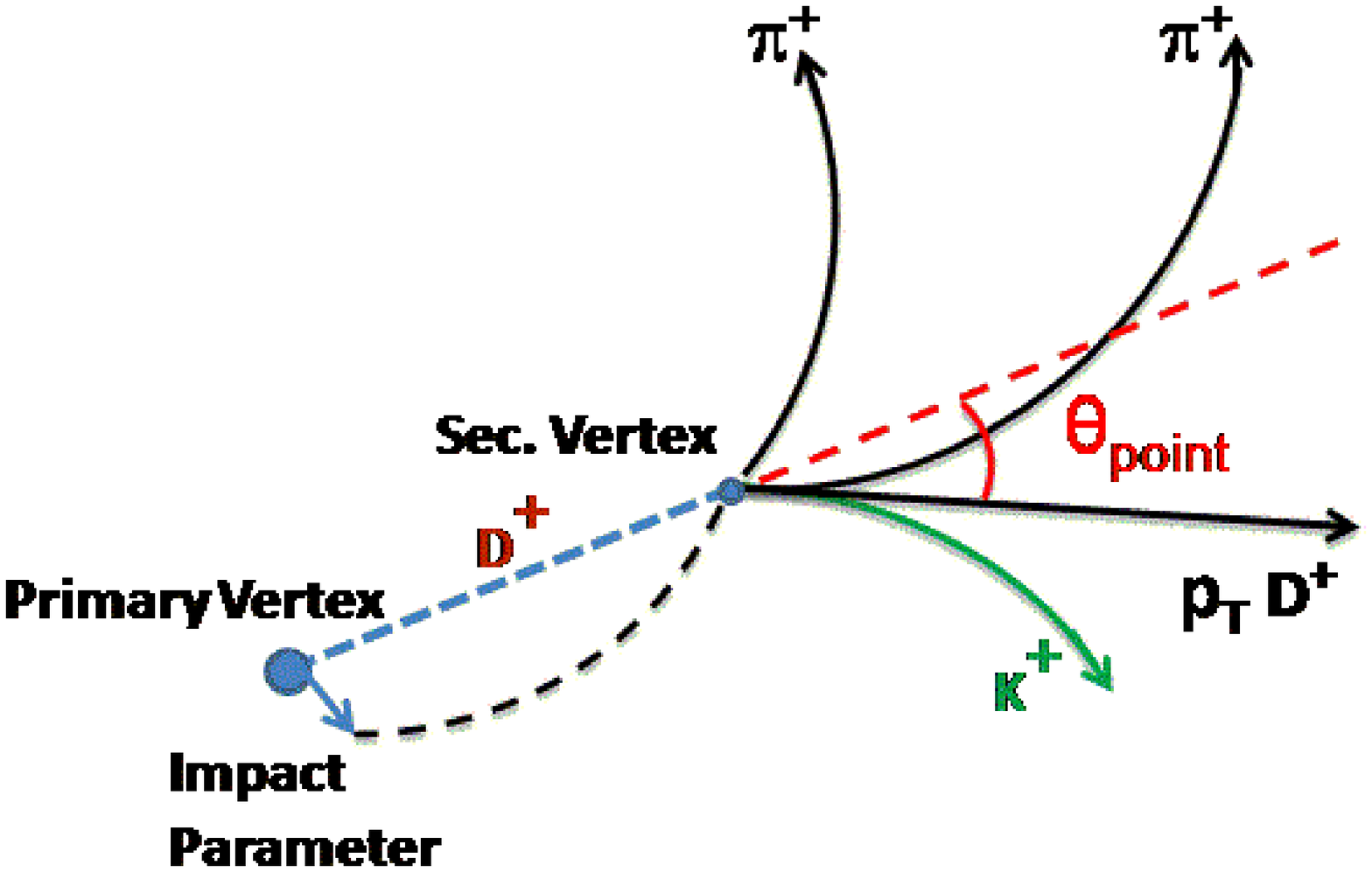}
   \hspace{18mm}
   \includegraphics[width=16pc,height=10pc]{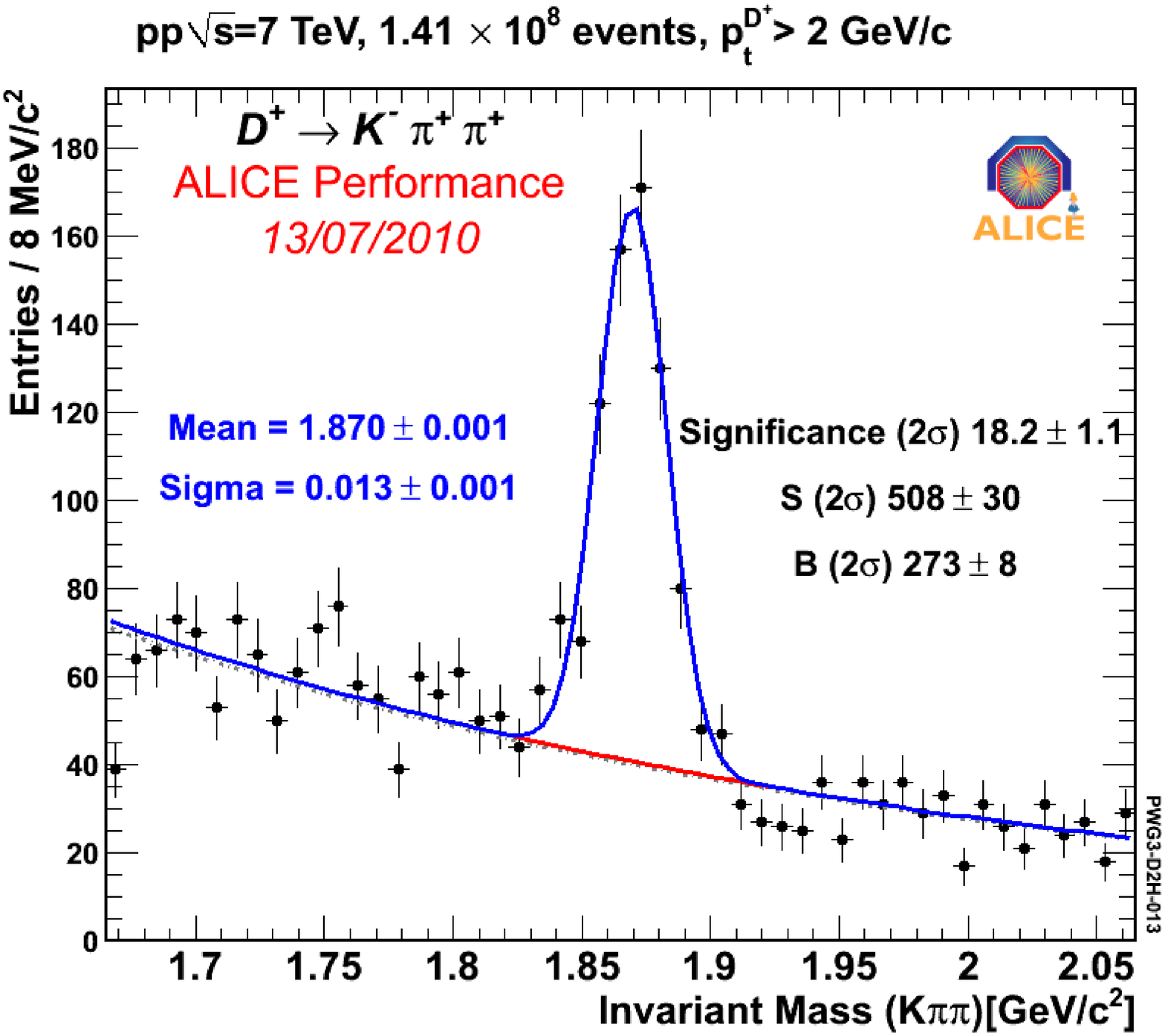}
   \caption{Sketch of $D^{+}$ decay topology (left hand panel) and $K^{-}
     \pi^{+} \pi^{+}$ invariant mass spectra for $p_{t} > 2$ GeV/c  (right hand panel) }
   \label{dplus}
 \end{figure*}
 Fig \ref{dplus} (lower panel) shows the invariant mass spectra for $D^{+}$ obtained
 after applying the topological cuts. With $10^{9}$ pp
$events^{1}$, we expect the $p_{t}$ integrated significance of 49 and larger than 10
for  $p_{t}$ up to $\sim$  10 GeV/c. 

\section{Expected sensitivity for the comparison  to pQCD prediction
  in pp collisions}
\begin{figure}
  \includegraphics[width=15pc,height=9pc]{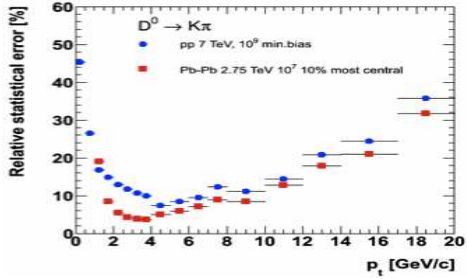}

\caption{Expected relative statistical error in 1 year of data taking
  for $D^{0} \rightarrow K^{-} \pi^{+}$.}
\label{sens}
\end{figure}

Fig. \ref{sens} shows the expected relative statistical error on the measured
$D^{0} $  distribution for p-p
collisions at 7 TeV ($10^{9}$ minimum bias pp events expected to
collect in 2010)
and central Pb-Pb collisions at 2.75 TeV ($10^{7}$ events expected to
collect in 2010 ). 

The accessible $p_{t}$ range is 1-20 GeV/c
for Pb-Pb and 0.5-20 GeV/c for p-p, with a point-by-point statistical
error less than 15-20$\%$. The systematic error (acceptance and
 efficiency correction, centrality selection for Pb-Pb) is expected to
 be smaller than 20$\%$.
 For the case of pp collisions, the experimental errors on the $p_{t}$
 differential cross section are expected to be significantly smaller
 than the current theoretical uncertainty from perturbative QCD
 calculations. 

\begin{figure}[!htbp]
\includegraphics[width=15pc,height=9pc]{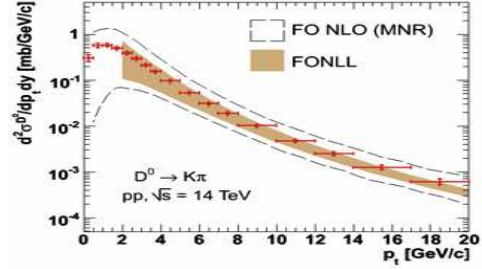}

\caption{Differential cross section of $D^{0}$ in pp at 14 TeV
  compared to NLO pQCD predictions from MNR and FONLL calculations.}
\label{cross}
\end{figure}

 In fig. \ref{cross}, we superimpose the simulated ALICE measurement
 points to the prediction band from the MNR fixed order massive
 calculations and from the FONLL fixed order
 next-to-leading log calculations \cite{fonll} for $D^{0}$ (same for $D^{+}$) at 14 TeV. The perturbative uncertainty band
 were estimated by varying the values of charm quark mass and of the
 factorization and re-normalization scales. The comparison
 \footnote{Here results are shown for 14 TeV but also expected a good
   sensitivity at 7 TeV} shows that ALICE has an excellent capabilty to perform a sensitive test of
 the pQCD predictions for charm production at LHC energy.

\section{Charm Energy Loss in Pb-Pb collisions: Nuclear Modification factor}
  The measured spectra in p-p and Pb-Pb can be used to compute the
  nuclear modification factor, $R_{AA}(p_t) = \frac{d^2 N_{AA}/dp_t
    dy}{<N_{coll}> d^2N_{pp}/dp_t dy}$. This observable is supposed to
  be 1 if the nucleus-nucleus collision behaves as a simple
  superposition of independent nucleon-nucleon collisions (Binary Scaling).
\begin{figure}[!htbp]
\includegraphics[width=15pc,height=9pc]{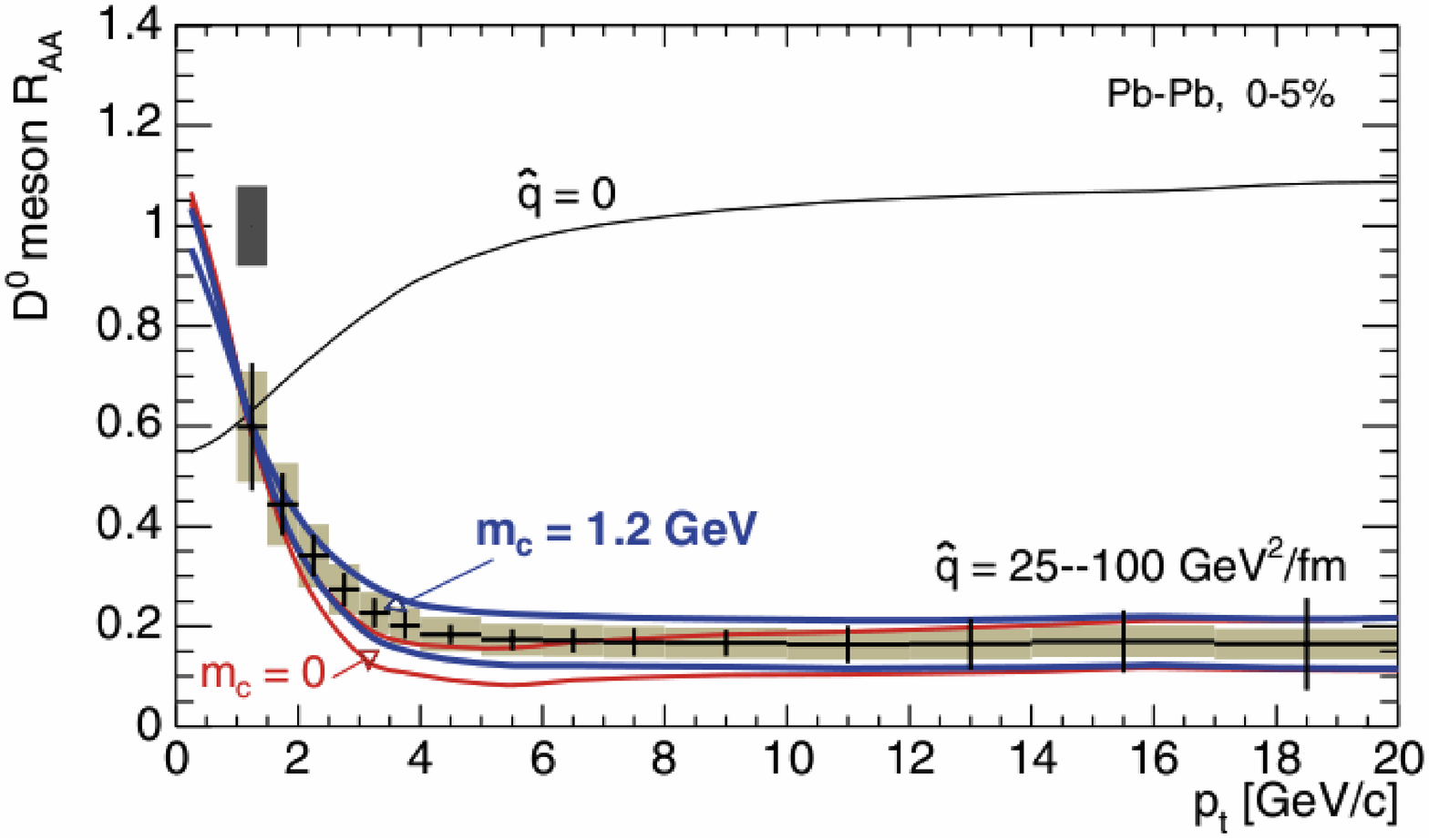}
\includegraphics[width=15pc,height=9pc]{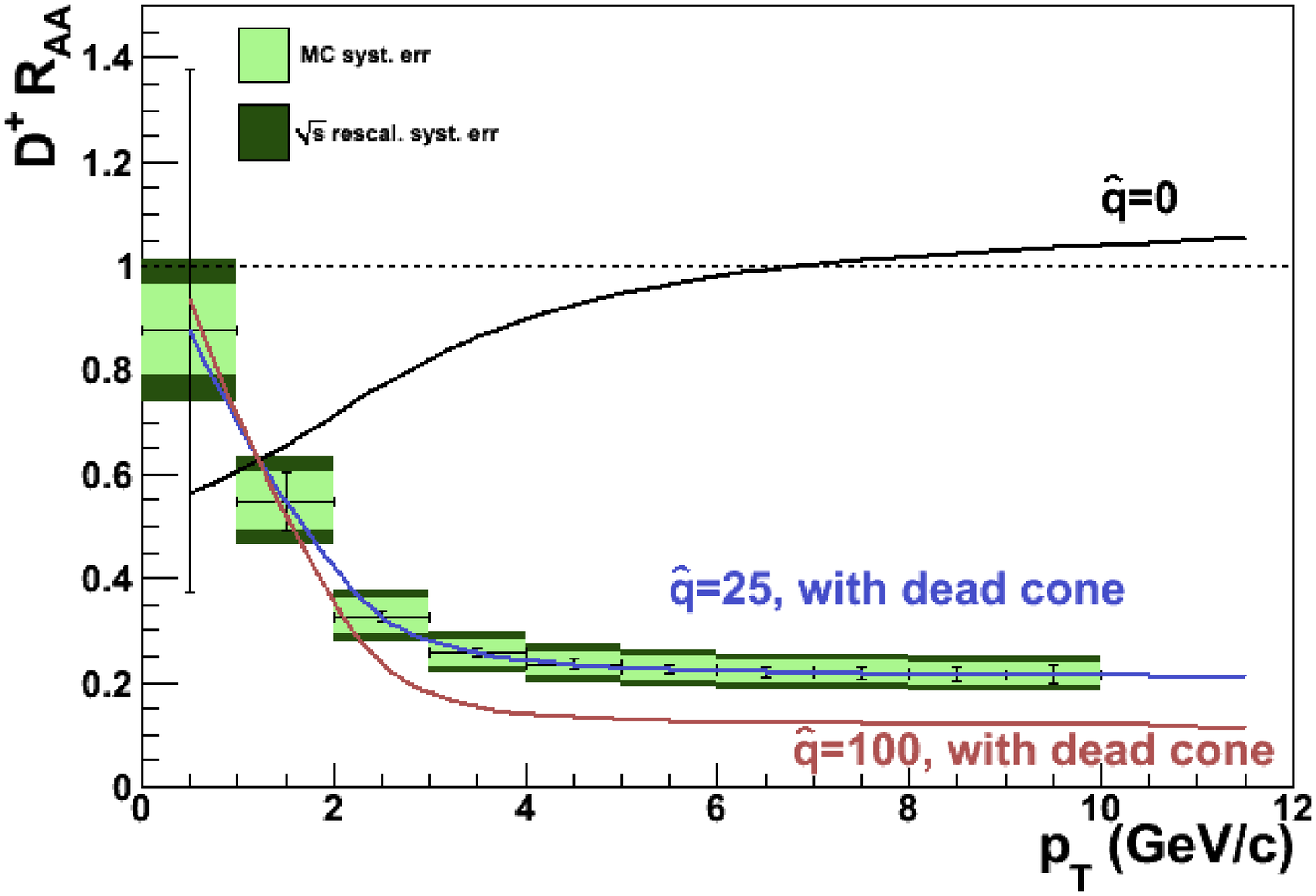}
\caption{\it Expected performance for the measurement of Nuclear
  Modification factor of $D^0$ mesons (upper panel) and $D^+$ (lower panel) after one year of data taking at nominal luminosity.}
  \label{energy_loss}
\end{figure}

The expected performance for the measurement of the nuclear modification factor for $D^0$ and $D^+$ mesons after one year of data taking  at nominal luminosity is shown in figure~\ref{energy_loss}. Theoretical calculations for different energy loss scenarios depending on the in-medium transport coefficient $\hat{q}$ and on the c-quark mass are also shown.  The bands corresponding to $m_c$ =1.2 GeV and $\hat{q}$= 25-100 $GeV^2$/fm reflects the estimated uncertainty on the model expectations for $R_{AA}^D$. The small difference between the two bands ($m_c$=0 and $m_c$=1.2 GeV) indicates that with respect to energy loss, charm behaves similarly to light quarks. Therefore, the enhancement of the heavy to light ratio is a sensitive measurement, essentially free of mass effects, to study the colour charge dependence of parton energy loss\cite{energy}.

\section{Conclusions}
The ALICE detector provides excellent tracking, vertexing and particle
identification to allow a high precision measurement of the open charm
cross section via hadronic decays, both in pp and Pb-Pb collisions and
over a wide range of transverse momenta.  We have shown the very first
results for the 2 hadronic decay channels ($D^{0} \rightarrow K^{-}
\pi^{+}$ and $D^{+}\rightarrow K^{-} \pi^{+} \pi^{+}$) with 1.4 $\times$ 10 $^{8}$
minimum bias pp events and also shown that with 10$^{9}$ pp events, these measurements
will provide sensitive tests for pQCD at LHC energy. The
corresponding measurements in Pb-Pb at 5.5 A TeV energy  will allows us to
measure energy loss with good precision.

\end{document}